\documentclass{ifacconf}
\usepackage[sort]{natbib}
\usepackage{eufrak}
\usepackage{url}

\usepackage{mysty}

\begin{document}
\begin{frontmatter}

\title{Higher-Dimensional Timed Automata\thanksref{spons}}

\thanks[spons]{This research is supported by the \textit{Chaire ISC~:
    Engineering Complex Systems} -- \smash{\'E}cole polytechnique --
  Thales -- FX -- DGA -- Dassault Aviation -- DCNS Research -- ENSTA
  ParisTech -- T{\'e}l{\'e}com ParisTech}

\author{Uli Fahrenberg}

\address{LIX, \smash{\'E}cole polytechnique, Palaiseau, France}

\begin{abstract}
  We introduce a new formalism of higher-dimensional timed automata,
  based on van~Glabbeek's higher-dimensional automata and Alur's timed
  automata.  We prove that their reachability is PSPACE-complete and
  can be decided using zone-based algorithms.  We also show how to use
  tensor products to combat state-space explosion and how to extend
  the setting to higher-dimensional hybrid automata.
\end{abstract}


\end{frontmatter}

\section{Introduction}

In approaches to non-interleaving concurrency, more than one event may
happen concurrently.  There is a plethora of formalisms for modeling
and analyzing such concurrent systems, \eg~Petri
nets~\citep{book/Petri62}, event
structures~\citep{DBLP:journals/tcs/NielsenPW81}, configuration
structures~\citep{DBLP:conf/lics/GlabbeekP95,
  DBLP:journals/tcs/GlabbeekP09},
or more recent variations such as dynamic event
structures~\citep{DBLP:conf/forte/ArbachKPN15} and Unravel
nets~\citep{conf/acmsac/CasuP17}.  They all share the convention of
differentiating between concurrent and interleaving executions; using
CCS notation~\citep{book/Milner89}, $a| b\ne a. b+ b. a$.

For modeling and analyzing embedded or cyber-physical systems,
formalisms which use real time are available.  These include timed
automata~\citep{DBLP:journals/tcs/AlurD94}, time Petri
nets~\citep{journal/transcom/MerlinF76}, timed-arc Petri
nets~\citep{DBLP:conf/apn/Hanisch93}, or various classes of hybrid
automata~\citep{DBLP:journals/tcs/AlurCHHHNOSY95}.  Common for them all
is that they identify concurrent and interleaving executions; here,
$a| b= a. b+ b. a$.

We are interested in formalisms for real-time non-interleaving
concurrency.  Hence we would like to differentiate between concurrent
and interleaving executions and be able to model and analyze real-time
properties.  Few such formalisms seem to be available in the
literature.  (The situation is perhaps best epitomized by the fact
that there is a natural non-interleaving semantics for Petri
nets~\citep{DBLP:journals/iandc/GoltzR83} which is also used in
practice~\citep{DBLP:conf/spin/Esparza10,
  DBLP:series/eatcs/EsparzaH08}, but almost all work on real-time
extensions of Petri nets~\citep{journal/transcom/MerlinF76,
  DBLP:conf/apn/Hanisch93, DBLP:conf/performance/Sifakis77,
  DBLP:conf/formats/Srba08}, including the popular tool
TAPAAL\footnotemark
, use an interleaving semantics.

Also Uppaal\footnotemark
, the successful tool for modeling and analyzing networks of timed
automata, uses an interleaving semantics for such networks.  This
leads to great trouble with state-space explosion (see also
Sect.~\ref{se:tensor} of this paper) which, we believe, can be avoided
with a non-interleaving semantics such as we propose here.

We introduce higher-dimensional timed automata (HDTA), a formalism
based on the (non-interleaving) higher-dimensional automata
of~\citet{DBLP:journals/tcs/Glabbeek06, Glabbeek91-hda}
and~\citet{DBLP:conf/popl/Pratt91} and the timed automata
of~\citet{DBLP:journals/tcs/AlurD94, DBLP:conf/icalp/AlurD90}.  We
show that HDTA can model interesting phenomena which cannot be
captured by neither of the formalisms on which they are based, but
that their analysis remains just as accessible as the one of timed
automata.  That is, reachability for HDTA is PSPACE-complete and can
be decided using zone-based algorithms.

In the above-mentioned interleaving real-time formalisms, continuous
flows and discrete actions are orthogonal in the sense that executions
alternate between real-time delays and discrete actions which are
immediate, \ie~take no time.  (In the hybrid setting, these are
usually called flows and mode changes, respectively.)
Already~\citet{DBLP:conf/stacs/SifakisY96} notice that this
significantly simplifies the semantics of such systems and hints that
this is a main reason for the success of these formalisms (see the
more recent~\citet{DBLP:conf/formats/Srba08} for a similar
statement).%
\footnotetext{\url{http://www.tapaal.net/}}%
\footnotetext{\url{http://www.uppaal.org/}}

In the (untimed) non-interleaving setting, on the other hand, events
have a (logical, otherwise unspecified) duration.  This can be seen,
for example, in the ST-traces of~\citet{DBLP:journals/tcs/Glabbeek06}
where actions have a start ($a^+$) and a termination ($a^-$) and are
(implicitly) running between their start and termination, or in the
representation of concurrent systems as Chu spaces over
$3=\{ 0, \frac12, 1\}$, where $0$ is interpreted as ``before'',
$\smash[t]{\frac12}$ as ``during'', and $1$ as ``after'',
see~\citet{DBLP:journals/mscs/Pratt00}.  Intuitively, only if events
have duration can one make statements such as ``while $a$ is running,
$b$ starts, and then while $b$ is running, $a$ terminates''.

In our non-interleaving real-time setting, we hence abandon the
assumption that actions are immediate.  Instead, we take the view that
actions start and then run during some \emph{specific} time before
terminating.  While this runs counter to the standard assumption in
most of real-time and hybrid modeling, a similar view can be found,
for example, in~\citet{DBLP:conf/icalp/Cardelli82}.\footnote{The
  author wishes to thank Kim G.~Larsen for pointing him towards this
  paper.}

Given that we abandon the orthogonality between continuous flows and
discrete actions, we find it remarkable to see that the standard
techniques used for timed automata transfer to our non-interleaving
setting.  Equally remarkable is, perhaps, the fact that even though
\textit{``[t]he timed-automata model is at the very border of
  decidability, in the sense that even small additions to the
  formalism [\dots] will soon lead to the undecidability of
  reachability questions''}~\citep{book/AcetoILS07}, our extension to
higher dimensions and non-interleaving concurrency is completely free
of such trouble.

The contributions of this paper are, thus, (1)~the introduction of a
new formalism of HDTA, a natural extension of higher-dimensional
automata and timed automata, in Sect.~\ref{se:hdta}; (2)~the proof
that reachability for HDTA is PSPACE-complete and decidable using
zone-based algorithms, in Sects.~\ref{se:reach} and~\ref{se:zone};
(3)~the introduction of a tensor product for HDTA which can be used
for parallel composition, in Sect.~\ref{se:tensor}; and (4)~the
extension of the definition to higher-dimensional hybrid automata
together with a non-trivial example of two independently bouncing
balls, in Sect.~\ref{se:hybrid}.

\section{Preliminaries}
\label{se:hda}

We recall a few facts about higher-dimensional automata and timed
automata.

\subsection{Higher-Dimensional Automata}

Higher-dimensional automata are a generalization of finite automata
which permit the specification of independence of actions through
higher-dimensional elements.  That is, they consist of states and
transitions, but also squares which signify that two events are
independent, cubes which denote independence of three events, \etc.
To introduce them properly, we need to start with precubical sets.

A \emph{precubical set} is a graded set $X= \bigcup_{ n\in \Nat} X_n$,
with $X_n\cap X_m= \emptyset$ for $n\ne m$, together with mappings
$\delta_{ k, n}^\nu:X_n\to X_{ n- 1}$, $k= 1,\dots, n$, $\nu= 0, 1$,
satisfying the \emph{precubical identity}
\begin{equation*}
  \delta_{ k, n- 1}^\nu \delta_{ \ell, n}^\mu= \delta_{ \ell- 1, n- 1}^\mu
  \delta_{ k, n}^\nu \qquad( k< \ell)\,.
\end{equation*}

\begin{figure}[b]
  \centering
  \begin{tikzpicture}[>=stealth', x=1cm, y=.8cm]
    \tikzstyle{every node}=[font=\footnotesize]
    \tikzstyle{every state}=[fill=white,shape=circle,inner
    sep=.5mm,minimum size=3mm]
    \path[fill=black!15] (0,0) to (2,0) to (2,2) to (0,2) to (0,0);
    \node[state] (00) at (0,0) {};
    \node[state] (10) at (2,0) {};
    \node[state] (01) at (0,2) {};
    \node[state] (11) at (2,2) {};
    \path (00) edge (01);
    \path (00) edge (10);
    \path (01) edge (11);
    \path (10) edge (11);
    \node at (1,1) {$x$};
    \node at (-.4,1.05) {$\delta_1^0 x$};
    \node at (2.4,1.05) {$\delta_1^1 x$};
    \node at (1,-.25) {$\delta_2^0 x$};
    \node at (1,2.25) {$\delta_2^1 x$};
    \node at (-1,-.35) {$\delta_1^0 \delta_2^0 x= \delta_1^0
      \delta_1^0 x$};
    \node at (-1,2.35) {$\delta_1^0 \delta_2^1 x= \delta_1^1
      \delta_1^0 x$};
    \node at (3,-.35) {$\delta_1^1 \delta_2^0 x= \delta_1^0
      \delta_1^1 x$};
    \node at (3,2.35) {$\delta_1^1 \delta_2^1 x= \delta_1^1
      \delta_1^1 x$};
  \end{tikzpicture}
  \caption{%
    \label{fi:2cubefaces-full}
    A $2$-cube $x$ with its four faces $\delta_1^0 x$, $\delta_1^1 x$,
    $\delta_2^0 x$, $\delta_2^1 x$ and four corners}
\end{figure}
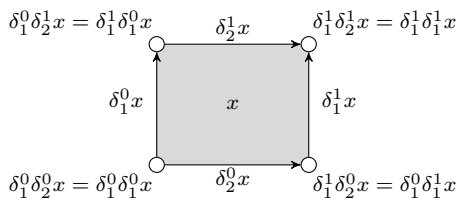

Elements of $X_n$ are called \emph{$n$-cubes}, and for $x\in X_n$,
$n= \dim x$ is its \emph{dimension}.  The mappings
$\delta_{ k, n}^\nu$ are called \emph{face maps}, and we will usually
omit the extra subscript $n$ and write $\delta_k^\nu$ instead of
$\delta_{ k, n}^\nu$.  Intuitively, each $n$-cube $x\in X_n$ has $n$
\emph{lower faces} $\smash[t]{\delta_1^0 x,\dotsc, \delta_n^0 x}$ and $n$
\emph{upper faces} $\delta_1^1 x,\dotsc, \delta_n^1 x$, and the
precubical identity expresses the fact that $( n- 1)$-faces of an
$n$-cube meet in common $( n- 2)$-faces; see
Fig.~\ref{fi:2cubefaces-full} for an example.

A precubical set $X$ is \emph{finite} if $X$ is finite as a set.  This
means that $X_n$ is finite for each $n\in \Nat$ and that $X$ is
\emph{finite-dimensional}: there exists $N\in \Nat$ such that
$X_n= \emptyset$ for all $n\ge N$.


Let $\Sigma$ be a finite set of \emph{actions} and recall that a
\emph{multiset} over $\Sigma$ is a mapping $\Sigma\to \Nat$.  We
denote multisets using double braces $\mlbrace\cdot\mrbrace$ and the
set of multisets over $\Sigma$ by $\Nat^\Sigma$.  The
\emph{cardinality} of $S\subseteq \Nat^\Sigma$ is
$| S|= \sum_{ a\in \Sigma} S( a)$.

A \emph{higher-dimensional automaton} (HDA) is a structure
$( X, x^0, X^f, \lambda)$, where $X$ is a finite precubical set with
initial state $x^0\in X_0$ and accepting states $X^f\subseteq X_0$,
and $\lambda: X\to \Nat^\Sigma$ is a labeling function such that for
every $x\in X$,
\begin{itemize}
\item $| \lambda( x)|= \dim x$,
\item $\lambda( \delta_k^0 x)= \lambda( \delta_k^1 x)$ for all
  $k\le n$, and
\item $\lambda( x)\setminus \lambda( \delta_k^0 x)$ is a singleton
  for all $k\le \dim x$.
\end{itemize}
The conditions on the labeling ensure that the label of an $n$-cube is
an extension, by one event, of the label of any of its faces.  The
computational intuition is that when passing from a lower face
$\delta_k^0 x$ of $x\in X$ to $x$ itself, the (unique) event in
$\lambda( x)\setminus \lambda( \delta_k^0 x)$ is started, and when
passing from $x$ to an upper face $\delta_\ell^1 x$, the event in
$\lambda( x)\setminus \lambda( \delta_\ell^1 x)$ is terminated.

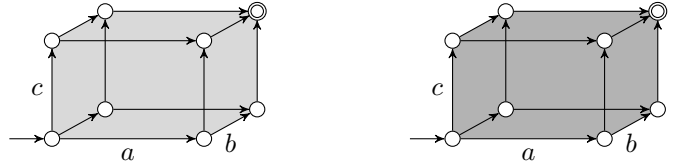
\begin{figure}[t]
  \centering
  \begin{tikzpicture}[->,>=stealth',auto,x=1cm,y=1.3cm]
    \begin{scope}
      \path[fill=black!15] (0,0) to (2,0) to (2.7,.3) to (2.7,1.3) to
      (.7,1.3) to (0,1);
      \node[state, initial] (000) at (0,0) {};
      \node[state] (001) at (0,1) {};
      \node[state] (010) at (.7,.3) {};
      \node[state] (011) at (.7,1.3) {};
      \node[state] (100) at (2,0) {};
      \node[state] (101) at (2,1) {};
      \node[state] (110) at (2.7,.3) {};
      \node[state, accepting] (111) at (2.7,1.3) {};
      \path (000) edge node[below] {$a$} (100);
      \path (000) edge node[left] {$c$} (001);
      \path (000) edge (010);
      \path (001) edge (011);
      \path (001) edge (101);
      \path (010) edge (110);
      \path (010) edge (011);
      \path (100) edge node[below] {$b$} (110);
      \path (100) edge (101);
      \path (011) edge (111);
      \path (101) edge (111);
      \path (110) edge (111);
    \end{scope}
    \begin{scope}[xshift=15em]
      \path[fill=black!30] (0,0) to (2,0) to (2.7,.3) to (2.7,1.3) to
      (.7,1.3) to (0,1);
      \node[state, initial] (000) at (0,0) {};
      \node[state] (001) at (0,1) {};
      \node[state] (010) at (.7,.3) {};
      \node[state] (011) at (.7,1.3) {};
      \node[state] (100) at (2,0) {};
      \node[state] (101) at (2,1) {};
      \node[state] (110) at (2.7,.3) {};
      \node[state, accepting] (111) at (2.7,1.3) {};
      \path (000) edge node[below] {$a$} (100);
      \path (000) edge node[left] {$c$} (001);
      \path (000) edge (010);
      \path (001) edge (011);
      \path (001) edge (101);
      \path (010) edge (110);
      \path (010) edge (011);
      \path (100) edge node[below] {$b$} (110);
      \path (100) edge (101);
      \path (011) edge (111);
      \path (101) edge (111);
      \path (110) edge (111);
    \end{scope}
  \end{tikzpicture}
  \caption{%
    \label{fi:cube}
    Two example HDA.  Left, the hollow cube; right, the full cube}
\end{figure}

HDA can indeed model higher-order concurrency of actions.  As an
example, the hollow cube on the left of Fig.~\ref{fi:cube}, consisting
of all six faces of a cube but not of its interior, models the
situation where the actions $a$, $b$ and $c$ are mutually independent,
but cannot be executed all three concurrently.  The full cube on the
right of Fig.~\ref{fi:cube}, on the other hand, has $a$, $b$ and $c$
independent as a set.  The left HDA might model a system of three
users connected to two printers, so that every two of the users can
print concurrently but not all three, whereas the right HDA models a
system of three users connected to (at least) three printers.

\subsection{Timed Automata}

Timed automata extend finite automata with clock variables and
invariants which permit the modeling of real-time properties.
Let $C$ be a finite set of \emph{clocks}.  $\Phi( C)$ denotes the set
of \emph{clock constraints} defined as
\begin{multline*}
  \Phi( C)\ni \phi_1, \phi_2::= c\bowtie k\mid \phi_1\land
  \phi_2 \\
  ( c\in C, k\in \Int, \bowtie\in\{ \mathord<, \mathord\le,
  \mathord\ge, \mathord>\})\,.
\end{multline*}
Hence a clock constraint is a conjunction of comparisons of clocks to
integers.

A \emph{clock valuation} is a mapping $v: C\to \Realnn$, where
$\Realnn$ denotes the set of non-negative real numbers.  The
\emph{initial} clock valuation is $v^0: C\to \Realnn$ given by
$v^0( c)= 0$ for all $c\in C$.  For $v\in \Realnn^C$, $d\in \Realnn$,
and $C'\subseteq C$, the clock valuations $v+ d$ and $v[ C'\gets 0]$
are defined by
\begin{equation*}
  ( v+ d)( c)= v( c)+ d\,; \quad v[ C'\gets 0]( c)=
  \begin{cases}
    0 &\text{if } c\in C'\,, \\
    v( c) &\text{if } c\notin C'\,.
  \end{cases}
\end{equation*}
For $v\in \Realnn^C$ and $\phi\in \Phi( C)$, we write $v\models \phi$
if $v$ satisfies $\phi$ and
$\sem \phi=\{ v: C\to \Realnn\mid v\models \phi\}$.

A \emph{timed automaton} is a structure $( Q, q^0, Q^f, I, E)$, where
$Q$ is a finite set of locations with initial location $q^0\in Q$ and
accepting locations $Q^f\subseteq Q$, $I: Q\to \Phi( C)$ assigns
invariants to states, and
$E\subseteq Q\times \Phi( C)\times \Sigma\times 2^C\times Q$ is a set
of guarded transitions.

The \emph{semantics} of a timed automaton $A=( Q, q^0, Q^f, I, E)$ is
a (usually infinite) transition system
$\sem A=( S, s^0, S^f, \mathord\leadsto)$, with
$\mathord\leadsto\subseteq S\times S$, given as follows:
\begin{align*}
  S &=\{( q, v)\subseteq Q\times \Realnn^C\mid v\models I( q)\} \\
  s^0 &= ( l^0, v^0) \quad S^f= S\,\cap\, Q^f\!\!\times\! \Realnn^C \\
  \mathord\leadsto &= \{(( q, v),( q, v+ d))\mid \forall 0\le d'\le d:
  v+ d'\models I( q)\} \\
  &\quad \cup\{(( q, v),( q', v'))\mid \exists( q, \phi, a, C', q')\in
  E: \\
  &\hspace{12.5em} v\models \phi, v'= v[ C'\gets 0]\}
\end{align*}
Note that we are ignoring the labels here, as we will be concerned
with reachability only.  As usual, we say that $A$ is \emph{reachable}
iff there exists a finite path $s^0\leadsto\dotsm\leadsto s$ in $\sem
A$ for which $s\in S^f$.

The definition of $\leadsto$ ensures that actions are immediate:
whenever $( q, \phi, a, C', q')\in E$, then $A$ passes from $( q, v)$
to $( q', v')$ without any delay.  Time progresses only during delays
$( q, v)\leadsto( q, v+ d)$ in locations.

\section{Higher-Dimensional Timed Automata}
\label{se:hdta}

Unlike timed automata, higher-dimensional automata make no formal
distinction between states ($0$-cubes), transitions ($1$-cubes), and
higher-dimensional cubes.  We transfer this intuition to
higher-dimensional timed automata, so that each $n$-cube has an
invariant which specifies when it is enabled and an exit condition
giving the clocks to be reset when leaving:

\begin{defn}
  A \emph{higher-dimensional timed automaton} (HDTA) is a structure
  $( L, l^0, L^f, \lambda, \inv, \exit)$, where
  $( L, l^0, L^f, \lambda)$ is a finite higher-dimensional automaton
  and $\inv: L\to \Phi( C)$, $\exit: L\to 2^C$ assign \emph{invariant}
  and \emph{exit} conditions to each $n$-cube.
\end{defn}

The \emph{semantics} of a HDTA
$A=( L, l^0, L^f, \lambda, \inv, \exit)$ is a (usually infinite)
transition system $\sem A=( S, s^0, S^f, \mathord\leadsto)$, with
$\mathord\leadsto\subseteq S\times S$, given as follows:
\begin{align*}
  S &=\{( l, v)\subseteq L\times \Realnn^C\mid v\models \inv( l)\} \\
  s^0 &= ( l^0, v^0) \quad S^f= S\,\cap\, L^f\!\!\times\! \Realnn^C \\
  \mathord\leadsto &= \{(( l, v),( l, v+ d))\mid \forall 0\le d'\le d:
  v+ d'\models \inv( l)\} \\
  &\quad \cup\{(( \delta_k^0 l, v),( l, v'))\mid k\in\{ 1,\dotsc, \dim
  l\},  \\
  &\hspace{10em} v'= v[ \exit( \delta_k^0 l)\gets 0]\models \inv( l)\}
  \\
  &\quad \cup\{(( l, v),( \delta_k^1 l, v'))\mid k\in\{ 1,\dotsc, \dim
  l\}, \\
  &\hspace{10em} v'= v[ \exit( l)\gets 0]\models \inv( \delta_k^1 l)\}
\end{align*}
We omit labels from the semantics, as we will be
concerned only with \emph{reachability}:
%
Given a HDTA $A$, does there exist a finite path
$s^0\leadsto\dotsm\leadsto s$ in $\sem A$ such that $s\in S^f$?

Note that in the definition of $\leadsto$ above, we allow time to
evolve in any $n$-cube in $L$.  Hence transitions (\ie~$1$-cubes) are
not immediate.  The second line in the definition of $\leadsto$
defines the passing from an $( n-1)$-cube to an $n$-cube, \ie~the
start of a new concurrent event, and the third line describes what
happens when finishing a concurrent event.  Exit conditions specify
which clocks to reset when leaving a cube.

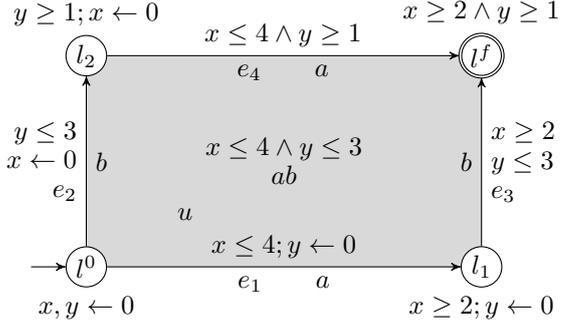
\begin{figure}[t]
  \centering
  \begin{tikzpicture}[>=stealth', x=1.3cm, y=.7cm]
    \begin{scope}
      \path[fill=black!15] (0,0) -- (4,0) -- (4,4) -- (0,4);
      \node[state, initial left] (00) at (0,0) {$l^0$};
      \node[state] (10) at (4,0) {$l_1$};
      \node[state] (01) at (0,4) {$l_2$};
      \node[state, accepting] (11) at (4,4) {$l^f$};
      \node[below] at (00.south) {$x, y\gets 0$};
      \node[below] at (10.south) {$x\ge 2; y\gets 0$};
      \node[above] at (01.north) {$y\ge 1; x\gets 0$};
      \node[above] at (11.north) {$x\ge 2\land y\ge 1$};
      \path (00) edge node[above] {$x\le 4; y\gets 0$} node[below]
      {$e_1 \qquad a$} (10);
      \path (00) edge node[left, align=right] {$y\le 3$ \\ $x\gets 0$
        \\ $e_2$} node[right] {$b$} (01);
      \path (10) edge node[right, align=left] {$x\ge 2$ \\$y\le 3$ \\
        $e_3$} node[left] {$b$} (11);
      \path (01) edge node[above] {$x\le 4\land y\ge 1$} node[below]
      {$e_4 \qquad a$} (11);
      \node[align=center] at (2,2) {$x\le 4\land y\le 3$ \\ $ab$};
      \node at (1,1) {$u$};
    \end{scope}
  \end{tikzpicture}
  \caption{%
    \label{fi:thda-ex1}
    The HDTA of Example~\ref{ex:thds-ex1}}
\end{figure}

\begin{exmp}
  \label{ex:thds-ex1}
  We give a few examples of two-dimensional timed automata.  The
  first, in Fig.~\ref{fi:thda-ex1}, models two actions, $a$ and $b$,
  which can be performed concurrently.  It consists of four states
  ($0$-cubes) $l^0, l_1, l_2, l^f$, four transitions ($1$-cubes) $e_1$
  through $e_4$, and one $ab$-labeled square ($2$-cube)~$u$.
  This HDTA models that performing $a$ takes between two and four time
  units, whereas performing $b$ takes between one and three time
  units.  To this end, we use two clocks $x$ and $y$ which are reset
  when the respective actions are started and then keep track of how
  long they are running.

  Hence $\exit( l^0)=\{ x, y\}$, and the invariants $x\le 4$ at the
  $a$-labeled transitions $e_1$, $e_4$ and at the square $u$ ensure
  that $a$ takes at most four time units.  The invariants $x\ge 2$ at
  $l_1$, $e_3$ and $l^f$ take care that $a$ cannot finish before two
  time units have passed.  Note that $x$ is also reset when exiting
  $e_2$ and $l_2$, ensuring that regardless when $a$ is started,
  whether before $b$, while $b$ is running, or after $b$ is
  terminated, it must take between two and four time units.
\end{exmp}

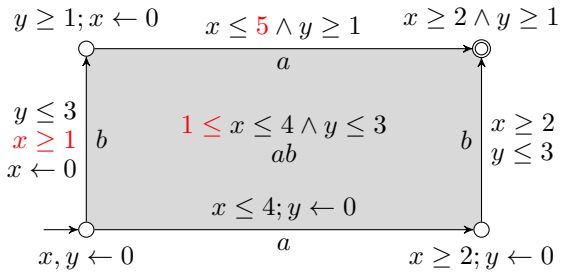
\begin{figure}[b]
  \centering
  \begin{tikzpicture}[>=stealth', x=1.3cm, y=.6cm]
    \begin{scope}[yshift=-5cm]
      \path[fill=black!15] (0,0) -- (4,0) -- (4,4) -- (0,4);
      \node[state, initial left] (00) at (0,0) {};
      \node[state] (10) at (4,0) {};
      \node[state] (01) at (0,4) {};
      \node[state, accepting] (11) at (4,4) {};
      \node[below] at (00.south) {$x, y\gets 0$};
      \node[below] at (10.south) {$x\ge 2; y\gets 0$};
      \node[above] at (01.north) {$y\ge 1; x\gets 0$};
      \node[above] at (11.north) {$x\ge 2\land y\ge 1$};
      \path (00) edge node[above] {$x\le 4; y\gets 0$} node[below]
      {$a$} (10);
      \path (00) edge node[left, align=right] {$y\le 3$ \\
        $\textcolor{red}{x\ge 1}$ \\ $x\gets 0$} node[right] {$b$}
      (01);
      \path (10) edge node[right, align=left] {$x\ge 2$ \\ $y\le 3$}
      node[left] {$b$} (11);
      \path (01) edge node[above] {$x\le \textcolor{red}{5}\land y\ge
        1$} node[below]
      {$a$} (11);
      \node[align=center] at (2,2) {$\textcolor{red}{1\le{}} x\le
        4\land y\le 3$ \\ $ab$};
    \end{scope}
  \end{tikzpicture}
  \caption{%
    \label{fi:thda-ex2}
    The HDTA of Example~\ref{ex:thda-ex2}}
\end{figure}

\begin{exmp}
  \label{ex:thda-ex2}
  In the HDTA shown in Fig.~\ref{fi:thda-ex2} (where we have omitted
  the names of states etc.\ for clarity and show changes to
  Fig.~\ref{fi:thda-ex1} in \textcolor{red}{red}), invariants have
  been modified so that $b$ can only start after $a$ has been running
  for one time unit, and if $b$ finishes before $a$, then $a$ may run
  one time unit longer.  Hence an invariant $x\ge 1$ is added to the
  two $b$-labeled transitions and to the $ab$-square (at the
  right-most $b$-transition $x\ge 1$ is already implied), and the
  condition on $x$ at the top $a$-transition is changed to $x\le 5$.
  Note that the left edge is now permanently disabled: before entering
  it, $x$ is reset to zero, but its edge invariant is $x\ge 1$.  This
  is as expected, as $b$ should not be able to start before~$a$.
\end{exmp}

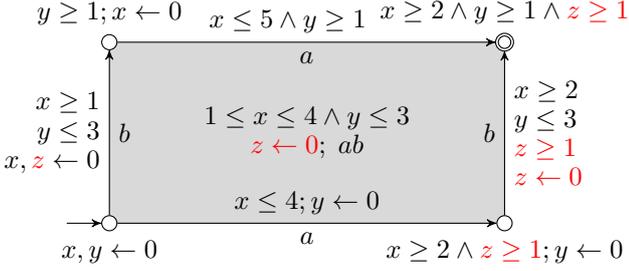
\begin{figure}[tbp]
  \centering
  \begin{tikzpicture}[>=stealth', x=1.3cm, y=.6cm]
    \begin{scope}[yshift=-10cm]
      \path[fill=black!15] (0,0) -- (4,0) -- (4,4) -- (0,4);
      \node[state, initial left] (00) at (0,0) {};
      \node[state] (10) at (4,0) {};
      \node[state] (01) at (0,4) {};
      \node[state, accepting] (11) at (4,4) {};
      \node[below] at (00.south) {$x, y\gets 0$};
      \node[below] at (10.south) {$x\ge 2\land \textcolor{red}{z\ge 1};
        y\gets 0$};
      \node[above] at (01.north) {$y\ge 1; x\gets 0$};
      \node[above] at (11.north) {$x\ge 2\land y\ge 1\land
        \textcolor{red}{z\ge 1}$};
      \path (00) edge node[above] {$x\le 4; y\gets 0$} node[below]
      {$a$} (10);
      \path (00) edge node[left, align=right] {$x\ge 1$ \\ $y\le
        3$ \\ $x, \textcolor{red}{z}\gets 0$} node[right]
      {$b$} (01);
      \path (10) edge node[right, align=left] {$x\ge 2$ \\ $y\le 3$ \\
        $\textcolor{red}{z\ge 1}$ \\ $\textcolor{red}{z\gets 0}$}
      node[left] {$b$} (11);
      \path (01) edge node[above, pos=.45] {$x\le 5\land y\ge
        1$} node[below]
      {$a$} (11);
      \node[align=center] at (2,2) {$1\le x\le 4\land y\le 3$ \\
        $\textcolor{red}{z\gets 0};\; ab$};
    \end{scope}
  \end{tikzpicture}
  \caption{%
    \label{fi:thda-ex3}
    The HDTA of Example~\ref{ex:thda-ex3}}
\end{figure}

\begin{exmp}
  \label{ex:thda-ex3}
  The HDTA in Fig.~\ref{fi:thda-ex3} (where we show changes to
  Fig.~\ref{fi:thda-ex2} in \textcolor{red}{red}) models the
  additional constraint that $b$ also \emph{finish} one time unit
  before $a$.  To this end, an extra clock $z$ is introduced which is
  reset when $b$ terminates and must be at least $1$ when $a$ is
  terminating.  After these changes, the right $b$-labeled edge is
  deadlocked: when leaving it, $z$ is reset to zero but needs to be at
  least one when entering the accepting state.  Again, this is
  expected, as $a$ should not terminate before~$b$.

  As both vertical edges are now permanently disabled, the accepting
  state can only be reached through the square.  This shows that
  reachability for HDTA cannot be reduced to one-dimensional
  reachability along transitions and relates them to the partial HDA
  of~\citet{DBLP:conf/calco/FahrenbergL15}.
\end{exmp}

\section{One-Dimensional Timed Automata}

We work out the relation between one-dimensional HDTA (\ie~1DTA) and
standard timed automata.  Note that this is not trivial, as in timed
automata, clocks can only be reset at transitions, and, semantically,
transitions take no time.  In contrast, in our 1DTA, resets can occur
in states and transitions may take time.

\begin{prop}
  There is a linear-time algorithm which, given any timed
  automaton $A$, constructs a 1DTA $A'$, with one extra clock, so that
  $A$ is reachable iff $A'$ is.
\end{prop}

\begin{pf}
  Let $A=( Q, q^0, Q^f, I, E)$ be a timed automaton.  It is clear that
  $L= Q\cup E$ forms a one-dimensional precubical set, with $L_0= Q$,
  $L_1= E$, $\delta_1^0( q, \phi, a, C', q')= q$, and
  $\delta_1^1( q, \phi, a, C', q')= q'$.  Let $l^0= q^0$ and
  $L^f= Q^f$.  In order to make transitions immediate, we introduce a
  fresh clock $c\notin C$.
  For $q\in Q$, let $\lambda( q)= \emptyset$, $\inv( q)= I( q)$, and
  $\exit( q)=\{ c\}$.  For $e=( q, \phi, a, C', q')\in E$, put
  $\lambda( e)=\{ a\}$, $\inv( e)= \phi\land( c\le 0)$, and
  $\exit( e)= C'$.  We have defined a 1DTA
  $A'=( L, l^0, L^f, \lambda, \inv, \exit)$ (over clocks
  $ C\cup\{ c\}$).
  As $c$ is reset whenever exiting a state, and every transition has
  $c\le 0$ as part of its invariant, it is clear that transitions in
  $A'$ take no time, and the claim follows.
\qed\end{pf}

\begin{prop}
  There is a linear-time algorithm which, given any 1DTA $A$,
  constructs a timed automaton $A'$ over the same clocks such that $A$
  is reachable iff $A'$ is.
\end{prop}

\begin{figure}[tbp]
  \centering
  \begin{tikzpicture}[>=stealth']
    \begin{scope}
      \node[state] (0) at (0,0) {};
      \node[state] (2) at (4,0) {};
      \node at (0,.3) {$\phi_1, C_1$};
      \node at (4,.3) {$\phi_3, C_3$};
      \path (0) edge node[above] {$\phi_2, C_2$} node[below] {$a$}
      (2);
    \end{scope}
    \path (2,-.5) edge[double] (2,-1.2);
    \begin{scope}[yshift=-1.9cm]
      \node[state] (0) at (0,0) {};
      \node[state] (1) at (2,0) {};
      \node[state] (2) at (4,0) {};
      \node at (0,.3) {$\phi_1$};
      \node at (2,.3) {$\phi_2$};
      \node at (4,.3) {$\phi_3$};
      \path (0) edge node[above] {$\ltrue, C_1$} (1);
      \path (1) edge node[above] {$\ltrue, C_2$} node[below] {$a$}
      (2);
    \end{scope}
  \end{tikzpicture}
  \caption{%
    \label{fi:ta-to-t1da}
    Conversion of 1DTA edge to timed automaton}
\end{figure}
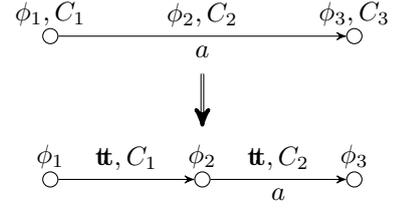

\begin{pf}
  Let $A=( L, l^0, L^f, \lambda, \inv, \exit)$ be a 1DTA, we construct
  a timed automaton $A'=( Q, q^0, Q^f, I, E)$.  Because transitions in
  $A$ may take time, we cannot simply let $Q= L_0$, but need to add
  extra states corresponding to the edges in $L_1$.  Let, thus,
  $Q= L$, $I= \inv$, and
  $E=\{( \delta_1^0 x, \ltrue, \tau, \exit( \delta_1^0 x), x) ,( x,
  \ltrue, \lambda( x), \exit( x), \delta_1^1 x)\mid x\in L_1\}$,
  where $\tau\notin \Sigma$ is a fresh (silent) action.  See
  Fig.~\ref{fi:ta-to-t1da}.
\qed\end{pf}

Note that even though silent transitions in timed automata are a
delicate matter~\citep{DBLP:journals/fuin/BerardPDG98}, the fact that
we add them in the last proof is unimportant as we are only concerned
with reachability.  PSPACE-completeness of reachability for timed
automata now implies the following:

\begin{cor}
  \label{co:PSPACE}
  Reachability for HDTA is PSPACE-hard.
\end{cor}

\section{Reachability for HDTA is in PSPACE}
\label{se:reach}

We now turn to extend the notion of \emph{regions} to HDTA, in order
to show that reachability for HDTA is decidable in PSPACE.

\begin{defn}
  Let $( L, l^0, L^f, \lambda, \inv, \exit)$ be a HDTA and
  $R\subseteq L\times \Realnn^C\times L\times \Realnn^C$.  Then $R$ is
  an \emph{untimed bisimulation} if
  $(( l^0, v^0),( l^0, v^0))\in R$ and, for all
  $(( l_1, v_1),( l_2, v_2))\in R$,
  \begin{compactitem}
  \item $l_1\in L^f$ iff $l_2\in L^f$;
  \item whenever $( l_1, v_1)\leadsto( l_1', v_1')$, then also $( l_2,
    v_2)\leadsto( l_2', v_2')$ for some $(( l_1', v_1'),( l_2',
    v_2'))\in R$;
  \item whenever $( l_2, v_2)\leadsto( l_2', v_2')$, then also $( l_1,
    v_1)\leadsto( l_1', v_1')$ for some $(( l_1', v_1'),( l_2',
    v_2'))\in R$.
  \end{compactitem}
\end{defn}

For a HDTA $A$, let $\cmax A$ denote the maximal constant appearing in
any $\inv( l)$ for $l\in L$, and let $\req$ denote standard region
equivalence~\citep{DBLP:journals/tcs/AlurD94}.  Extend $\req$ to
$\sem A$ by defining $( l, v)\req( l', v')$ iff $l= l'$ and
$v\req v'$.

\begin{lem}
  $\req$ is an untimed bisimulation.
\end{lem}

\begin{pf}
  This follows from standard properties of region
  equivalence~\citep{DBLP:journals/tcs/AlurD94}.
\qed\end{pf}





For any HDTA $A$, the \emph{quotient} of
$\sem A=( S, s^0, S^f, \mathord\leadsto)$ under an untimed
bisimulation $R$ is defined, as usual, as
$\sem A/ R=( S/ R,[ s^0]_R, S^f/ R, \mathord{ \tilde\leadsto})$, where
$S/ R$ is the set of equivalence classes, $[ s^0]_R$ is the
equivalence class in which $s^0$ belongs, and
$\mathord{ \tilde\leadsto}\subseteq S/ R\times S/ R$ is defined by
$\tilde s\mathrel{ \tilde\leadsto} \tilde s'$ iff $\exists s\in \tilde
s, s'\in \tilde s': s\leadsto s'$.

\begin{lem}
  Let $A$ be a HDTA and $R$ an untimed bisimulation on $A$.
  Then $A$ is reachable iff $\sem A/ R$ is.
\end{lem}

\begin{pf}
  By definition, an accepting location is reachable in $A$ iff an
  accepting state is reachable in $\sem A$.  On $\sem A$, $R$ is a
  standard bisimulation, hence the claim follows.
\qed\end{pf}

\begin{lem}
  For any HDTA $A$, the quotient $\sem A/ \mathord{ \req}$ is finite.
\end{lem}

\begin{pf}
  This follows immediately from the standard fact that the set of
  clock regions, \ie~$\Realnn^C/ \mathord{ \req}$, is
  finite~\citep{DBLP:journals/tcs/AlurD94}.
\qed\end{pf}

The size of $\sem A/ \mathord{ \req}$ is exponential in the size of
$A$, but reachability in $\sem A/ \mathord{ \req}$ can be decided in
PSPACE, see~\citep{DBLP:journals/tcs/AlurD94}.  Together with
Corollary~\ref{co:PSPACE}, we conclude:

\begin{thm}
  Reachability for HDTA is PSPACE-complete.
\end{thm}

\section{Zone-Based Reachability}
\label{se:zone}

We show that the standard zone-based algorithm for checking
reachability in timed automata also applies in our HDTA setting.  This
is important, as zone-based reachability checking is at the basis of
the success of tools such as Uppaal,
see~\citep{DBLP:journals/sttt/LarsenPY97}.

Recall that the set $\Phi^+(C)$ of \emph{extended clock constraints}
over $C$ is defined by the grammar
\begin{multline*}
  \Phi^+( C)\ni \phi_1, \phi_2::= c\bowtie k\mid c_1- c_2\bowtie
  k\mid \phi_1\land \phi_2 \\
  ( c, c_1, c_2\in C, k\in \Int, \bowtie\in\{ \mathord<, \mathord\le,
  \mathord\ge, \mathord>\}),
\end{multline*}
and that a \emph{zone} over $C$ is a subset $Z\subseteq \Realnn^C$
which can be represented by an extended clock constraint $\phi$,
\ie~such that $Z= \sem \phi$.  Let $\mcal Z( C)$ denote the set of
zones over $C$.

For a zone $Z\in \mcal Z( C)$ and $C'\subseteq C$, the \emph{delay}
and \emph{reset} of $Z$ are given by $Z^\uparrow=\{ v+ d\mid v\in Z\}$
and $Z[ C'\gets 0]=\{ v[ C'\gets 0]\mid v\in Z\}$; these are again
zones, and their representation by an extended clock constraint can be
efficiently computed~\citep{DBLP:conf/ac/BengtssonY03}.  Also zone
inclusion $Z'\subseteq Z$ can be efficiently decided.

The \emph{zone graph} of a HDTA
$A=( L, l^0, L^f, \lambda, \inv, \exit)$ is a (usually infinite)
transition system $Z( A)=( S, s^0, S^f, \mathord\leadsto)$, with
$\mathord\leadsto\subseteq S\times S$, given as follows:
\begin{align*}
  S &=\{( l, Z)\subseteq L\times \mcal Z( C)\mid Z\subseteq \sem{
    \inv( l)}\} \\
  s^0 &= ( l^0, \sem{ v^0}^\uparrow\cap\sem{ \inv( l^0)}) \qquad S^f=
  S\,\cap\, L^f\!\!\times\! \mcal Z( C) \\
  \mathord\leadsto &= \{(( \delta_k^0 l, Z),( l, Z'))\mid k\in\{
  1,\dotsc, \dim l\}, \\
  &\hspace{9.5em} Z'= Z[ \exit( \delta_k^0 l)\gets 0]^\uparrow\cap
  \sem{ \inv( l)}\} \\
  &\;\cup\{(( l, Z),( \delta_k^1 l, Z'))\mid k\in\{ 1,\dotsc, \dim
  l\}, \\
  &\hspace{9.5em} Z'= Z[ \exit( l)\gets 0]^\uparrow\cap \sem{ \inv(
    \delta_k^1 l)}\}
\end{align*}

\begin{lem}
  For any HDTA $A$, an accepting location is reachable in $A$ iff an
  accepting state is reachable in $Z( A)$.
\end{lem}

\begin{pf}
  This follows from standard arguments as to the soundness and
  completeness of the zone abstraction~\citep{DBLP:journals/tcs/AlurD94}.
\qed\end{pf}

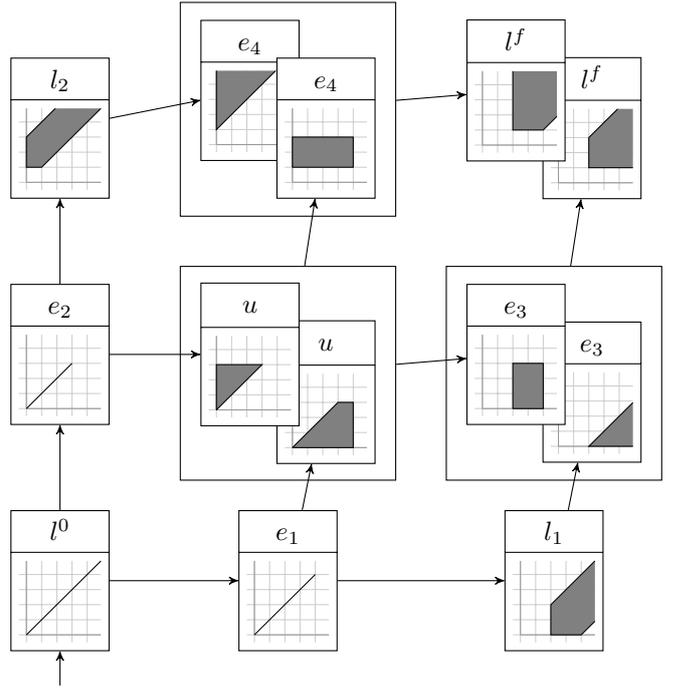
\begin{figure}[t]
  \centering
  \begin{tikzpicture}[>=stealth']
    \node[state with output, initial below] (l0) at (0,0)
    {$\vphantom{e_1}l^0$ \nodepart{two}%
      \begin{tikzpicture}[scale=.2,>=]
        \draw[gray!40, very thin] (-.5,-.5) grid (4.9,4.9);
        \draw[gray!80, thin] (0,0) to (4.9,0);
        \draw[gray!80, thin] (0,0) to (0,4.9);
        \draw (0,0) -- (4.9,4.9);
      \end{tikzpicture}
    };
    \node[state with output] (e1) at (3,0) {$\vphantom{l^0}e_1$
      \nodepart{two}%
      \begin{tikzpicture}[scale=.2,>=]
        \draw[gray!40, very thin] (-.5,-.5) grid (4.9,4.9);
        \draw[gray!80, thin] (0,0) to (4.9,0);
        \draw[gray!80, thin] (0,0) to (0,4.9);
        \draw (0,0) -- (4,4);
      \end{tikzpicture}
    };
    \node[state with output] (l1) at (6.5,0) {$\vphantom{l^0}l_1$
      \nodepart{two}%
      \begin{tikzpicture}[scale=.2,>=]
        \draw[gray!40, very thin] (-.5,-.5) grid (4.9,4.9);
        \draw[gray!80, thin] (0,0) to (4.9,0);
        \draw[gray!80, thin] (0,0) to (0,4.9);
        \draw[fill=gray] (4.9,4.9) -- (2,2) -- (2,0) -- (4,0) --
        (4.9,.9);
      \end{tikzpicture}
    };
    \node[state with output] (u1) at (3.5,2.5) {$\vphantom{l^0_1}u$
      \nodepart{two}%
      \begin{tikzpicture}[scale=.2,>=]
        \draw[gray!40, very thin] (-.5,-.5) grid (4.9,4.9);
        \draw[gray!80, thin] (0,0) to (4.9,0);
        \draw[gray!80, thin] (0,0) to (0,4.9);
        \draw[fill=gray] (0,0) -- (4,0) -- (4,3) -- (3,3) -- (0,0);
      \end{tikzpicture}
    };
    \node[state with output] (e2) at (0,3) {$\vphantom{l^0}e_2$
      \nodepart{two}%
      \begin{tikzpicture}[scale=.2,>=]
        \draw[gray!40, very thin] (-.5,-.5) grid (4.9,4.9);
        \draw[gray!80, thin] (0,0) to (4.9,0);
        \draw[gray!80, thin] (0,0) to (0,4.9);
        \draw (0,0) -- (3,3);
      \end{tikzpicture}
    };
    \node[state with output] (l2) at (0,6) {$\vphantom{l^0}l_2$
      \nodepart{two}%
      \begin{tikzpicture}[scale=.2,>=]
        \draw[gray!40, very thin] (-.5,-.5) grid (4.9,4.9);
        \draw[gray!80, thin] (0,0) to (4.9,0);
        \draw[gray!80, thin] (0,0) to (0,4.9);
        \draw[fill=gray] (1.9,4.9) -- (0,3) -- (0,1) -- (1,1) --
        (4.9,4.9);
      \end{tikzpicture}
    };
    \node[state with output] (u2) at (2.5,3) {$\vphantom{l^0_1}u$
      \nodepart{two}%
      \begin{tikzpicture}[scale=.2,>=]
        \draw[gray!40, very thin] (-.5,-.5) grid (4.9,4.9);
        \draw[gray!80, thin] (0,0) to (4.9,0);
        \draw[gray!80, thin] (0,0) to (0,4.9);
        \draw[fill=gray] (0,0) -- (0,3) -- (3,3) -- (0,0);
      \end{tikzpicture}
    };
    \node[draw, rectangle] (u) at (3,2.75) {%
      \begin{tikzpicture}[scale=2.6,>=]
        \path (0,0) -- (1,0) -- (1,1) -- (0,1) -- (0,0);
      \end{tikzpicture}
    };
    \node[state with output] (e41) at (2.5,6.5) {$\vphantom{l^0}e_4$
      \nodepart{two}%
      \begin{tikzpicture}[scale=.2,>=]
        \draw[gray!40, very thin] (-.5,-.5) grid (4.9,4.9);
        \draw[gray!80, thin] (0,0) to (4.9,0);
        \draw[gray!80, thin] (0,0) to (0,4.9);
        \draw[fill=gray] (3.9,4.9) -- (0,1) -- (0,4.9);
      \end{tikzpicture}
    };
    \node[state with output] (e42) at (3.5,6) {$\vphantom{l^0}e_4$
      \nodepart{two}%
      \begin{tikzpicture}[scale=.2,>=]
        \draw[gray!40, very thin] (-.5,-.5) grid (4.9,4.9);
        \draw[gray!80, thin] (0,0) to (4.9,0);
        \draw[gray!80, thin] (0,0) to (0,4.9);
        \draw[fill=gray] (0,1) -- (4,1) -- (4,3) -- (0,3) -- (0,1);
      \end{tikzpicture}
    };
    \node[draw, rectangle] (e4) at (3,6.25) {%
      \begin{tikzpicture}[scale=2.6,>=]
        \path (0,0) -- (1,0) -- (1,1) -- (0,1) -- (0,0);
      \end{tikzpicture}
    };
    \node[state with output] (e31) at (7,2.5) {$\vphantom{l^0}e_3$
      \nodepart{two}%
      \begin{tikzpicture}[scale=.2,>=]
        \draw[gray!40, very thin] (-.5,-.5) grid (4.9,4.9);
        \draw[gray!80, thin] (0,0) to (4.9,0);
        \draw[gray!80, thin] (0,0) to (0,4.9);
        \draw[fill=gray] (4.9,0) -- (2,0) -- (4.9,2.9);
      \end{tikzpicture}
    };
    \node[state with output] (e32) at (6,3) {$\vphantom{l^0}e_3$
      \nodepart{two}%
      \begin{tikzpicture}[scale=.2,>=]
        \draw[gray!40, very thin] (-.5,-.5) grid (4.9,4.9);
        \draw[gray!80, thin] (0,0) to (4.9,0);
        \draw[gray!80, thin] (0,0) to (0,4.9);
        \draw[fill=gray] (2,0) -- (4,0) -- (4,3) -- (2,3) -- (2,0);
      \end{tikzpicture}
    };
    \node[draw, rectangle] (e3) at (6.5,2.75) {%
      \begin{tikzpicture}[scale=2.6,>=]
        \path (0,0) -- (1,0) -- (1,1) -- (0,1) -- (0,0);
      \end{tikzpicture}
    };
    \node[state with output] (lf1) at (7,6)
    {$\vphantom{e_1}l^f$ \nodepart{two}%
      \begin{tikzpicture}[scale=.2,>=]
        \draw[gray!40, very thin] (-.5,-.5) grid (4.9,4.9);
        \draw[gray!80, thin] (0,0) to (4.9,0);
        \draw[gray!80, thin] (0,0) to (0,4.9);
        \path[fill=gray] (4.9,1) -- (2,1) -- (2,3) -- (3.9,4.9) --
        (4.9,4.9);
        \draw (4.9,1) -- (2,1) -- (2,3) -- (3.9,4.9);
      \end{tikzpicture}
    };
    \node[state with output] (lf2) at (6,6.5)
    {$\vphantom{e_1}l^f$ \nodepart{two}%
      \begin{tikzpicture}[scale=.2,>=]
        \draw[gray!40, very thin] (-.5,-.5) grid (4.9,4.9);
        \draw[gray!80, thin] (0,0) to (4.9,0);
        \draw[gray!80, thin] (0,0) to (0,4.9);
        \path[fill=gray] (4.9,1.9) -- (4,1) -- (2,1) -- (2,4.9) --
        (4.9,4.9);
        \draw (4.9,1.9) -- (4,1) -- (2,1) -- (2,4.9);
      \end{tikzpicture}
    };
    \path (l0) edge (e1);
    \path (e1) edge (l1);
    \path (e1) edge (u1);
    \path (l0) edge (e2);
    \path (e2) edge (l2);
    \path (e2) edge (u2);
    \path (l2) edge (e41);
    \path (u) edge (e42);
    \path (l1) edge (e31);
    \path (u) edge (e32);
    \path (e3) edge (lf1);
    \path (e4) edge (lf2);
  \end{tikzpicture}
  \caption{%
    \label{fi:zone}
    Zone graph of the HDTA in Fig.~\ref{fi:thda-ex1}}
\end{figure}

Any standard \emph{normalization}
technique~\citep{DBLP:conf/ac/BengtssonY03} may now be used to ensure
that the zone graph $Z( A)$ is finite, and then the standard zone
algorithms can be employed to efficiently decide reachability in HDTA.
As an example, Fig.~\ref{fi:zone} shows the zone graph of the HDTA in
Fig.~\ref{fi:thda-ex1} (Example~\ref{ex:thds-ex1}), with zones
displayed graphically using $x$ as the horizontal axis and $y$ as the
vertical.  (We have taken the liberty to simplify by computing unions
of zones at the locations $u$, $e_3$ and $e_4$ before proceeding.)

\section{Parallel Composition of HDTA}
\label{se:tensor}

There is a \emph{tensor product} on precubical sets which extends to
HDTA and can be used for parallel composition:

\begin{defn}
  Let $A_i=( L^i, l^{ i, 0}, L^{ i, f}, \lambda^i, \inv^i, \exit^i)$, for
  $i= 1, 2$, be HDTA.  The \emph{tensor product} of $A^1$ and $A^2$ is
  $A^1\otimes A^2=( L, l^0, L^f, \lambda, \inv, \exit)$ given as
  follows:
  \begin{gather*}
    L_n= \bigsqcup_{ p+ q= n} L^1_p\times L^2_q \quad\;\;
    l^0= ( l^{ 1, 0}, l^{ 2, 0}) \quad L^f= L^{ 1, f}\times L^{ 2,
      f} \\
    \delta_i^\nu( l^1, l^2)=
    \begin{cases}
      ( \delta_i^\nu l^1, l^2) &\!\!\!\!\text{if } i\le \dim l^1 \\
      ( l^1, \delta_{ i- \dim l^1}^\nu l^2) &\!\!\!\!\text{if } i>
      \dim l^1
    \end{cases} \\
    \lambda( l^1, l^2)= \lambda( l^1)\sqcup \lambda( l^2) \qquad
    \inv( l^1, l^2)= \inv( l^1)\land \inv( l^2) \\
    \exit( l^1, l^2)= \exit( l^1)\sqcup \exit( l^2)
  \end{gather*}
\end{defn}

Intuitively, tensor product is asynchronous parallel composition, or
independent product.  In combination with relabeling and restriction,
any parallel composition operator can be obtained,
see~\citet{WinskelN95-Models}
or~\citet{DBLP:conf/fossacs/Fahrenberg05} for the special case of HDA.

\begin{figure}[b]
  \centering
  \begin{tikzpicture}[>=stealth', x=.7cm, y=.3cm]
    \begin{scope}
      \node[state, initial below] (00) at (0,0) {};
      \node[state, accepting] (10) at (4,0) {};
      \node[state, initial below] (01) at (7,0) {};
      \node[state, accepting] (11) at (11,0) {};
      \node[above] at (00.north) {$x\gets 0$};
      \node[above] at (10.north) {$x\ge 2$};
      \node[above] at (01.north) {$y\gets 0$};
      \node[above] at (11.north) {$y\ge 1$};
      \path (00) edge node[above] {$x\le 4$} node[below] {$a$} (10);
      \path (01) edge node[above] {$y\le 3$} node[below] {$b$} (11);
    \end{scope}
  \end{tikzpicture}
  \caption{%
    \label{fi:tensor-ex}
    The two 1DTA of Example~\ref{ex:tensor-ex}}
\end{figure}
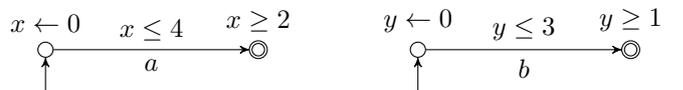

\begin{exmp}
  \label{ex:tensor-ex}
  Of the two 1DTA in Fig.~\ref{fi:tensor-ex}, the first models the
  constraint that performing the action $a$ takes between two and four
  time units, and the second, that performing $b$ takes between one
  and three time units.  (In the notation
  of~\citet{DBLP:conf/icalp/Cardelli82}, these are
  $a[2]\mathord: a(2)\mathord: 0$ and
  $b[1]\mathord: b(2)\mathord: 0$.)  Their tensor product is precisely
  the HDTA of Example~\ref{ex:thds-ex1}.
\end{exmp}

\begin{figure}[tbp]
  \centering
  \includegraphics*[width=.9\linewidth, bb=1.5cm 24cm 7cm 26.7cm]{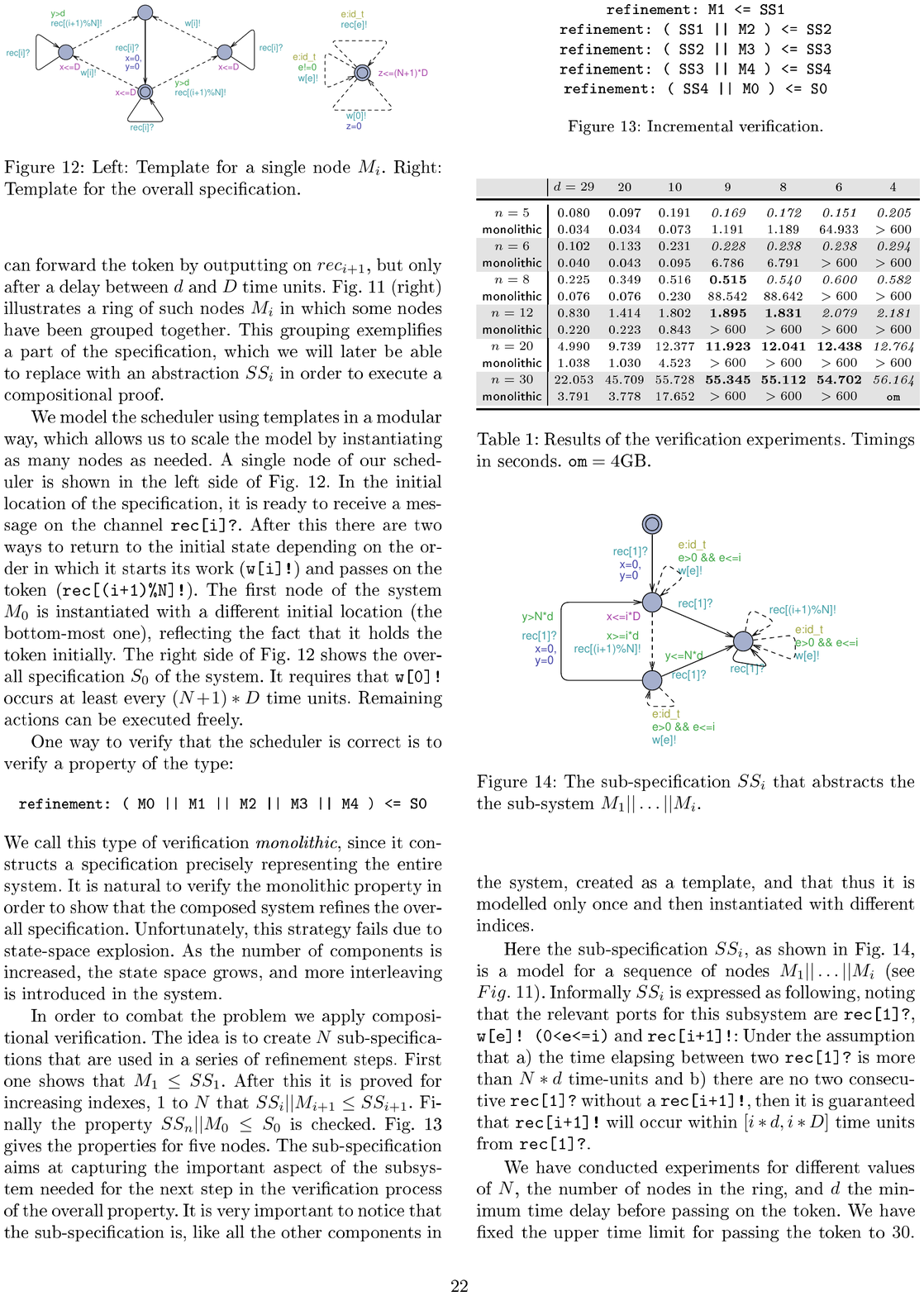}
  \caption{%
    \label{fi:milner}
    Uppaal model of a single node in Milner's scheduler}
\end{figure}

Using tensor product for parallel composition, one can avoid
introducing spurious interleavings and thus combat state-space
explosion.  Take the real-time version of Milner's scheduler
from~\citet{DBLP:journals/sttt/DavidLLNTW15} as an example.  This is
essentially a real-time round-robin scheduler in which the nodes are
simple timed automata, see Fig.~\ref{fi:milner} for the Uppaal model.

Notice that there are, essentially, two transitions from the initial
to the topmost state, one which outputs w[i] (``work'') and another
which passes on the token (rec[(i+1)\%N]!).  These transitions are
independent, but because of the limitations of the timed-automata
formalism, they have to be modeled as an interleaving diamond.  Thus,
when a number of such nodes ($N = 30$, say) are composed into a
scheduler, a high amount of interleaving is generated: but most of it
is spurious, owing to constraints of the modeling language rather than
properties of the system at hand.

\citet{DBLP:journals/sttt/DavidLLNTW15}~show that especially when $d$
is much smaller than $D$ (say, $d = 4$ and $D = 30$), verification of
the scheduler becomes impossible already for $N = 6$ nodes.  The
authors then show that this can be amended by using compositional
verification techniques, which are out of scope of our current work.

Another possibility is to use methods from partial order
reduction~\citep{DBLP:books/sp/Godefroid96} to \emph{detect} spurious
interleavings.  Aside from the fact that this has proven to be largely
impractical for timed automata, see for
example~\citet{DBLP:conf/cav/HansenLLN014}, we also argue that by
using HDTA as a modeling language, partial order reduction is, so to
speak, \emph{built into} the model.  Spurious interleavings are taken
care of during the modeling phase, instead of having to be detected
during the verification phase.

A 2DTA corresponding to the Uppaal model in Fig.~\ref{fi:milner} is
essentially the tensor product of the w[i] and rec[(i+1)\%N]!
transitions, together with $1$-dimensional loops and a transition from
the upper to lower corner for the rec[i]? input transitions.

\section{Higher-Dimensional Hybrid Automata}
\label{se:hybrid}

For completeness, we show that our definition of HDTA easily extends
to one for higher-dimensional hybrid automata.  Let $X$ be a finite
set of variables, $\dot X=\{ \dot x\mid x\in X$,
$X'=\{ x'\mid x\in X\}$, and $\Pred( Y)$ the set of (arithmetic)
predicates on free variables in $Y$.

\begin{defn}
  A \emph{higher-dimensional hybrid automaton} (HDHA) is a structure
  $( L, \lambda, \inv, \flow, \exit)$, where $( L, \lambda)$ is a
  finite higher-dimensional automaton and
  $\init, \inv: L\to \Pred( X)$,
  $\smash[t]{\flow: L\to \Pred( X\cup \dot X)}$, and
  $\exit: L\to \Pred( X\cup X')$ assign \emph{initial},
  \emph{invariant}, \emph{flow}, and \emph{exit} conditions to each
  $n$-cube.
\end{defn}

Note that we have removed initial and final locations from the
definition; this is standard for hybrid automata.

The \emph{semantics} of a HDHA
$A=( L, \lambda, \inv, \flow, \exit)$ is a (usually
infinite) transition system $\sem A=( S, S^0, \mathord\leadsto)$,
with $\mathord\leadsto\subseteq S\times S$, given as follows:
\begin{align*}
  S &= \{( l, v)\subseteq L\times \Realnn^X\mid v\models \inv( l)\} \\
  S^0 &= \{( l, v)\in S\mid v\models \init(l)\} \\
  \mathord\leadsto &= \{(( l, v),( l, v'))\mid \exists d\ge 0, f\in
  \Diff([ 0, d], \Real^X): \\
  &\hspace{8.2em} f( 0)= v, f( d)= v', \forall t\in \mathopen]
  0, d\mathclose[: \\
  &\hspace{8.2em} f( t)\models \inv( q),( f( t), \dot f( t))\models
  \flow( q)\} \\
  &\quad \cup\{(( \delta_k^0 l, v),( l, v'))\mid k\in\{ 1,\dotsc, \dim
  l\},  \\[-.5ex]
  &\hspace{14em} ( v, v')\models \exit( \delta_k^0 l)\}
  \\
  &\quad \cup\{(( l, v),( \delta_k^1 l, v'))\mid k\in\{ 1,\dotsc, \dim
  l\}, \\[-.5ex]
  &\hspace{14em} ( v, v')\models \exit( l)\}
\end{align*}
Here $\Diff(D_1, D_2)$ denotes the set of differentiable functions
$D_1\to D_2$.

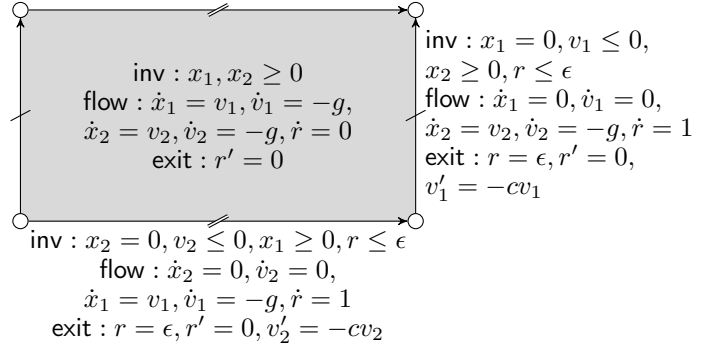
\begin{figure}[tbp]
  \centering
  \begin{tikzpicture}[>=stealth', x=1.3cm, y=.7cm]
    \begin{scope}
      \path[fill=black!15] (0,0) -- (4,0) -- (4,4) -- (0,4);
      \node[state] (00) at (0,0) {};
      \node[state] (10) at (4,0) {};
      \node[state] (01) at (0,4) {};
      \node[state] (11) at (4,4) {};
      \node[align=center] at (2,2) {%
        $\inv: x_1, x_2\ge 0$ \\
        $\flow: \dot x_1 = v_1, \dot v_1 = -g, 
        $
        \\
        $
        \dot x_2 = v_2, \dot v_2 = -g, \dot r = 0$ \\
        $\exit: r'= 0$
      };
      \path (10) edge node[right, align=left] {%
        $\inv: x_1 = 0, v_1\le 0,$ \\
        $x_2\ge 0, r\le \epsilon$ \\
        $\flow: \dot x_1 = 0, \dot v_1 = 0,$ \\
        $\dot x_2 = v_2, \dot v_2 = -g, \dot r = 1$ \\
        $\exit: r = \epsilon, r' = 0,$ \\
        $v_1' = -c v_1$
      } (11);
      \path (00) edge node[below, align=center] {%
        $\inv: x_2 = 0, v_2\le 0, x_1\ge 0, r\le \epsilon$ \\
        $\flow: \dot x_2 = 0, \dot v_2 = 0,$ \\
        $\dot x_1 = v_1, \dot v_1 = -g, \dot r = 1$ \\
        $\exit: r = \epsilon, r' = 0, v_2' = -c v_2$
      } (10);
      \path (00) edge (01);
      \path (01) edge (11);
      \path (-.1,1.9) edge[-] (.1,2.1);
      \path (3.9,1.9) edge[-] (4.1,2.1);
      \path (1.9, -.1) edge[-, double] (2.1,.1);
      \path (1.9, 3.9) edge[-, double] (2.1,4.1);
    \end{scope}
  \end{tikzpicture}
  \caption{%
    \label{fi:bouncing}
    Two independently bouncing balls}
\end{figure}

\begin{exmp}
  As a non-trivial example, we show a 2DHA which models two
  independently bouncing balls, following the \emph{temporal
    regularization} from~\citet{journals/scl/JohanssonELS99}, in
  Fig.~\ref{fi:bouncing}.  Here, the $2$-cube models the state in
  which both balls are in the air.  Its left and right edges are
  identified, as are its lower and upper edges, so that logically,
  this model is a torus.

  Its left / right edge is the state in which the second ball is in
  the air, whereas the first ball is in its $\epsilon$-regularized
  transition ($\epsilon> 0$) from falling to raising ($v_1' = -c v$,
  for some $c\in \mathopen] 0, 1\mathclose[$).  Similarly, its lower /
  upper edge is the state in which the first ball is in the air, while
  the second ball is $\epsilon$-transitioning.

  Due to the identifications, there is only one $0$-cube, which models
  the state in which both balls are $\epsilon$-transitioning; its
  $\inv$, $\flow$ and $\exit$ conditions can be inferred from the ones
  given.  With a notion of tensor product similar to the one for HDTA,
  this model can also be obtained as tensor product of the
  one-dimensional models for the individual balls.
\end{exmp}

\section{Conclusion}

We have seen that our new formalism of higher-dimensional timed
automata is useful for modeling interesting properties of
non-interleaving real-time systems, and that reachability for HDTA is
PSPACE-complete, but can be decided using zone-based algorithms.

We believe that our notion that in a non-interleaving real-time
setting, events should have a time duration, is quite natural.
Working on non-interleaving real-time semantics for Petri nets,
\citet{DBLP:conf/formats/ChatainJ13} remark that \textit{``[t]ime and
  causality [do] not necessarily blend well in [...] Petri nets''} and
propose to \emph{let time run backwards} to get nicer semantics.  We
should like to argue that our proposal of letting events have duration
appears more natural.

We have also seen how tensor product of HDTA can be used for parallel
composition, and that HDTA can easily be generalized to
higher-dimensional hybrid automata.  We believe that altogether, this
defines a powerful modeling formalism for non-interleaving real-time
systems. 

\bibliography{mybib}

\end{document}